\documentclass{article}
\usepackage{fancyvrb}
\usepackage{graphicx}
\begin{document}

\begin{Verbatim}[commandchars=\\\{\}]
                                                          88LA01T 
   DEUTERATED POLYSTYRENE-- SYNTHESIS AND USES FOR ULTRACOLD
            NEUTRON BOTTLES AND THE NEUTRON EDM EXPERIMENT

                                  by
                           Steve K. LAMOREAUX
   University of Washington Physics Dept., Seattle, WA 98195, USA
               and Institut Laue-Langevin, Grenoble, France

                               ABSTRACT
                         
The synthesis and application of deuterated polystyrene (dps) films
is discussed.  Ultracold neutron storage properties and the Fermi 
potential of dps films are measured  with the result that the storage 
lifetime is 700+/- 200 sec for dps in the bottle used  and the wall
(Fermi)  potential is about 165 neV.  The behavior under application 
of high electric fields in vacuum is measured; the films are sufficiently 
stable to use in a neutron EDM  bottle. Also, measurement of the 
relaxation rate of nuclear spin polarized 199Hg on dps films is 
measured giving a wall lifetime of 20 sec/cm mean free path, which 
should make the development of a 199Hg volume comagnetometer possible.




        1. Introduction................................2
        2. Synthesis of dps............................2
        3. Application of dps films....................3
        4. Neutron Loss Rate on dps....................4
        5. Measurement of dps wall (Fermi)
           potential...................................4
        6. High Voltage Vacuum 
           Characteristics.............................4
        7. 199Hg Nuclear Spin Polar- 
           ization Relaxation on dps...................5
        8. Conclusions.................................6

            Literature Cited

            Figures


\end{Verbatim}
\vfill
\eject

\section{Introduction}

Polystyrene is a plastic that is readily soluble in many organic solvents, particularly toluene. Such solutions are widely used industrially for their adhesive and coating properties. When applied to a variety of surfaces, upon evaporation of the solvent, they leave a
continuous film of polystyrene which adheres well to the substrate.
Surfaces on which the coating works well include fused silica, pyrex, and most metals. The chemical formulae of styrene and polystyrene are shown in Fig. 1.

Deuterated hydrocarbon coatings are useful for making ultracold neutron (UCN) storage bottles$^1$; 
for C$_8$D$_8$, the wall (Fermi) potential is 172 nev (calculated), which is comparable to stainless steel, and has relatively low total scattering and absorption cross sections. Deuterated polystyrene (dps), which is of this stoichiometric formula, has been demonstrated
to give excellent neutron storage lifetimes in addition to reasonable $^{199}$Hg nuclear spin polarization lifetimes (20 seconds per cm of mean free path), good vacuum properties, and  
high-voltage vacuum stability of thin films deposited on conducting media. For these reasons, we are considering dps for coating a UCN/$^{199}$ Hg neutron EDM bottle where nuclear spin polarized $^{199}$Hg will serve as a comagnetometer.

\section{Synthesis of dps}
 
Deuterated styrene (the monomer) is available from Sigma Chemicals; a convenient amount is 5g. ampoule, cat. no. S-7755, and is guaranteed 98+ atom\% D enrichment. A ``trace" of hydroquinone is added to prevent spontaneous polymerization and thus increase the shelf life. The easiest way to effect 
the polymerization is through heat and pressure;$^2$ the is contained in a sealed tube and heated to
180$^\circ$C, the details of which follow.
 
Fig. 2a is the plan of the reaction vessel used. The thick-walled Pyrex tube can withstand up to 10 bar pressure. The handle is included to facilitate manipulation of the tube during cooling and sealing. 

The d-styrene is transferred to the reaction tube by use of a pipette of design given in Fig. 2b. The tip of the pipette is thin to allow insertion through the neck of the reaction tube and to the bottom of the d-styrene ampoule. A rubber bulb should be used to fill the pipette and the transfer done under a fume hood as the styrene vapors are toxic and irritating.
 
Note that all of the glassware used should be very clean
and dry to keep the final product clean and to obtain 
reproducible results. A hot chromic acid bath followed by a 
deionized H$_2$0 rinse, then baking for an hour at 100$^\circ$C should be sufficient. 
The tubing used to make the reacticin vessel should be cleaned before fabrication to its final form. 

After the styrene is transferred to the reaction tube, the tube is dipped in liquid nitrogen to freeze the styrene, lowering its vapor pressure and reducing the risk of fire when the tube is sealed. Care should be taken to not let LN$_2$ into the tube itself (insert half-way) and the cooling should not be too rapid. Refer to Fig. 3. 

After the styrene is frozen and cooled to well below its melting point, the reaction tube is withdrawn from the LN$_2$ and an oxygen-gas flame is used to seal the neck. The freezing-sealing operation should be done as rapidly as possible.
 
When the reaction tube has warmed to a temperature which allows it to be handled, the tube is place in an oven and heated to 180$^\circ$C over a period of 30 min., held at 180$^\circ$C for 30 min., and then allowed to cool. The entire heating/cooling process is repeated three times.
 
The polymerization is now complete. The outside of the reaction tube is cleaned and wrapped in several layers of clean aluminum foil and the reaction tube is crushed (with
a hammer or vise). The glass shards are separated from the cylindrical ``stick" of dps with tweezers; any remaining glass fragments will settle to the bottom of the solvent container when the dps is dissolved. 

The dps is dissolved in deuterated toluene. The stick is simply dropped into a bottle of d-toluene and the dps dissolves over a period of about one day. The bottle should be amber glass with a polyethylene cap liner. The 25g size of d-toluene available from Sigma Chemicals (cat. no. T 2886) will admit the stick, however, the resulting solution has a viscosity about that of room temperature molasses. 
will admit the stick, however, the resulting solution has a viscosity about that of room temperature molasses. However, the solution can be used as a ``stock" to be added to d-toluene
as required to get the proper consistency for spraying, brushing, or dipping. The solution seems to be quite stable over time.

\section{Application of dps Films}

The dps film can be applied by dipping, brushing, or spraying.
So far only the first two methods have been used and have given identical results. The viscosity is adjusted as required for use in an airbrush or for painting; for dipping, the final thickness
depends on the viscosity.

The film is simply left to air dry. There is no evidence of deterioration (1 day) exposure to the atmosphere (water vapor, etc.). 

Dipping works well for small pieces; for larger pieces, probably spraying (helium pressurized air brush) will be most effective. Note that the application implements must be clean and dry.
 
A note of caution: Toulene vapors are quite irritating, particularly to the eyes. Prolonged exposure is known to cause conjuctivitis$^3$ and even short exposure to low vapor concentrations can cause discomfort. The symptoms include a ``gritty" sensation of the front of the eyeballs much like much like
``sunburned eyes," a problem skiers frequently encounter. The  
symptoms appear about six hours after exposure and last about 1 hours for a 20 min exposure to the vapor when working in an unventilated room. Toluene also has a narcotic: effect not unlike ether and breathing the vapor can make one quite giddy. 
Thus, all applications should be done in a fume hood; Eye protection and a respirator might be indicated, particularly when spraying.

\section{Neutron Loss Rate on dps Surface}

Measurement of the loss rate have been completed by measuring  
the UCN storage lifetime in a bottle with 20\% of the surface coated with dps and the rest with Fomblin grease. Figure 4 is a schematic of the experimental apparatus.  The reason only 20
 
The bottle UCN storage time was then measured with the entire surface coated with 
Fomblin grease and the loss rate due to the dps can be determined as follows:
$$\Gamma_{loss}=1/T_{store}$$
$$=f\Gamma_{Fomblin}+(1-f)\Gamma_{dps}+\Gamma_{\beta}$$
where $f$ is the fraction of the bottle area covered with Fomblin, 
$(1-f)$ is the fraction covered with dps, $\Gamma_\beta$ is the $\beta-$decay rate, and $T_{store}$ is the measured storage lifetime. Results for $f=.8$ and for $f=l$ are given in Table 1 and are plotted in
Fig. 5.  The results imply that $\Gamma_{dps} = 700\pm200$ seconds for the bottle geometry. 

This result is quite significant because for many surfaces with a high wall (Fermi) potential (stainless steel, for example) give a storage time of only about 20 seconds for bottles of a design similar to the one used for testing the dps.$^4$  The implication is that there is very little surface hydrogen on the dps film. It also implies that the bulk hydrogen ($<2$\%) has little effect. More work regarding the wall loss rate is needed, particularly a bottle with the entire surface coated with dps should be tested.

\section{Measurement of the dps wall (Fermi) potential} 
Refer to Fig. 6 for a schematic of the apparatus used. UCN obtained from PN5 are selected between a range of O and $\Delta E$ by D. Richardson's energy selector$^5$ and allowed to fall a distance $h$ meters to an aluminum foil coated with dps. The transmission of the foil is measured for $h=1.6$ m as a function of $\Delta E$. The neutrons gain kinetic energy of 100 neV/m fall with momentum in the direction necessary to penetrate the dps film.  Results indicate that the wall (Fermi) potential is $162\pm2$ neV 
in close agreement with the theoretical value 167 neV where the stoichiometric formula C$_8$D$_8$ and density 1.1g/cc was used, and scattering length .6674 and .6653 femtometers for carbon and deuterium, respectively.$^6$ There is some uncertainty in the density of the dps; the value used was obtained from the density of ordinary polystyrene by assuming the same molecular density and substituting D for H in the molecule. It is unlikely that the value obtained is accurate to more than 10\%. 

\section{High Voltage Vacuum Characteristics} 
Figure 7a is a schematic of the apparatus used to test the HT properties of organic films on metal surfaces. We have noticed with a number of fluorocarbon substances (Fomblin liquid, Fomblin grease, and Teflon) covering the HT electrodes, that when more than 10 kV is applied across the insulator (independent of electrode separation), the leakage current suddenly increases rapidly and the vacuum increases from $10^{-5}$ to more than $10^{-3}$ mbar. 
Apparently, the system is unstable; a small spark, perhaps from cosmic ray origin, decomposes some of the coating material; the discharge current increases causing more material to be decomposed further increasing the discharge current, etc. 

With dps, the ``runaway" is not so drastic and occurs with  50kV between the electrodes (4.4 cm separation which implies 11kV/cm field intensity). However, the test system breaks down in the same way without the dps coating on the
electrodes, so the maximum potential is still unknown. 

The difference between the fluorocarbon compounds and dps regarding their HT/vacuum characteristics could be due to the difference in molecular weight of the decomposition products 
(a factor of two lower for dps) and possibly a higher polymer/monomer binding energy. It has been reported that materials with a lower molecular weight are more stable in vacuum HT system.$^8$ It should also be noted that dps has a high volume resistivity; about $10^{14}\ \Omega$-cm.$^6$
 
Clearly, more work in regard to vacuum/HT characteristics of dps is needed, particularly how much voltage can be applied across the film and whether the maximum voltage is a function 
of electrode separation. Fig. 7b shows a setup which should be used. One problem with the tests conc:lucted so far is a lack of good electrical contact between the electrodes and insulating cylinder . If the dielectric film gets between the insulator and electrode (in the groove) and if there are voids, the three-dielectric problem exists. Also the leakage current and dielectric charging displacement current probably cause microdischarging between the electrodes and insulator in
regions of poor electrical contact. 

\section{$^{199}$Hg nuclear spin polarization relaxation on dps films}

Figure 8 shows a test cell design and the schematic of a setup used to test $^{199}$Hg spin relaxation rate on dps films. The inside of the bulb was coated with dps by injecting about 1 ml of the stock dps/toluene solution into the bulb using a stainless steel needle and a glass syringe. THe bulb was rotated and tilted until the entire inside was coated and then placed in a 100$^\circ$ oven with the stem vertical and pointing down, allowing the excess to run out. After 2 hours, all of the toluene evaporated and a thin, fairly even film of dps remained. The layer thickness was on the oreder of several hundred nanometers as indicated by light interference.  The 
dps on the stopcock was carefully removed with toluene and cotton tipped applicators. 
The stopcock was greased with Fomblin grease and attached to a vacuum system and pumped out to about $10^{-6}$ mbar. Provisions were include to admit Hg vapor 
into the cell and to optically pump the atoms. Refer to 9 for a complete description of Hg optical pumping. 

Figure 9 shows a plot of a free-precession signal which was obtained by polarizinq the atoms by having a magnetic field parallel to the propagation direction of the incident circulary polarized resonance light then suddenly rotating the field by 90$^\circ$ and the free precession signal monitored by the
effect of the nuclear spin polarization direction on the light absorption. The decay rate is a combination of the light pumping rate (proportional to intensity) and the wall relaxation rate. 
The extrapolated wall rate at zero light intensity is about 20 sec. The mean free path length is given 
by
$$\langle x\rangle =\frac{2a}{\pi}\int_{-\pi/2}^{\pi/2}\cos\theta d\theta= \frac{4}{\pi}a$$
where $a$ is the bulb radius. This implies that the dps wall lifetime is 20 sec per cm mean free path. bottle, the mean free path will be about 15 to 20 cm giving nuclear spin polarization lifetime of 300 to 400 sec. Note 
that the signals shown in Fig. 9 were obtained with natural 
$^{199}$Hg.
When isotopically pure $^{199}$Hg used, is the signal to noise will be 10 times greater. Also, the noise is mostly of shot origin; in the larger bottle a higher photon flux can be used which will give an even lower noise.

\section{Conclusions}

Deuterated polystyrene films appear to have excellent UCN storage properties, good vacuum characteristics, and good HT/vacuum behavior. In addition, the polarized $^{199}$Hg storage characteristics are excellent. These properties lead us to consider the use of dps for a neutron EDM storage bottle which includes polarized $^{199}$Hg as a comagnetometer. The properties of dps have not yet been fully investigated; the preliminary results are most promising.
 
dps has the added advantage that the films are easily applied and removed. The synthesis of dps is straightforward and the starting material is readily available.

I thank R. Golub, J.M. Pendlebury, and W. Mampe for useful discussions. I am indebted to D. Richardson who measured the wall (Fermi) potential of dps and assisted in the lifetime measurements. Y. Chibane assisted in the HT measurements. I also thank I. Kilvington for relentless encouragement and helpful discussions. R.  Morely, the University of Washington glassblower, fabricated the glass apparatus and assisted in developing the polymerization technique.

\section{Literature Cited} 
\bigskip
\noindent
1.  J.M. Pendlebury, University of Sussex, Private Communication (1987). 

\noindent
2.  John K. Stille, Introduction to Polymer Chemistry, (Wiley,New York, 1967). 

\noindent
3.  Merck Index 

\noindent
4.  P. Ageron, W. Mampe, and A.I. Kilvington, Z. Physik B, Condensed Matter {\bf 59}, 261 (1985). 

\noindent
5.  D. Richardson, Ph.D. Thesis, Sussex University (in preparation). 

\noindent
6.  L. Koestler and W.B. Yelon, Summary of Low Energy Neutron Scattering Lengths and Cross Sections. 

\noindent
7.  CRC Handbook of Chemistry and Physics. 

\noindent
8.  J.G. Trump and R.J. van de Graaff, Journal of Applied Physics {\bf 18}, 327  (1947). 

\noindent
9.  S.K. Lamoreaux, Ph.D. Thesis, University of Washington,Seattle, WA, USA (1986).

\section{Tables and Figures}

\begin{Verbatim}[commandchars=\\\{\}]

         TABLE 1--Results of Lifetime Measurement, time in Seconds 


     dps+Fomblin Grease  
     TIME   COUNTS(STD DEV)
         40   2310(37)  
         80   1986(36) 
         120  1742(15)  
         130  1712(50)  
         160  1511(52)  
         220  1278(45)  
         310    946(31)  

     Fomblin Grease  

         40   2387(61)  
         80   2005(43)  
         120  1693(88)  
         130  1694(38)  
         160  1512(42)  
         220  1253(32)  
         310   927(25)  

\end{Verbatim}
	\begin{centering}
	\begin{figure}
		\includegraphics[scale=1]{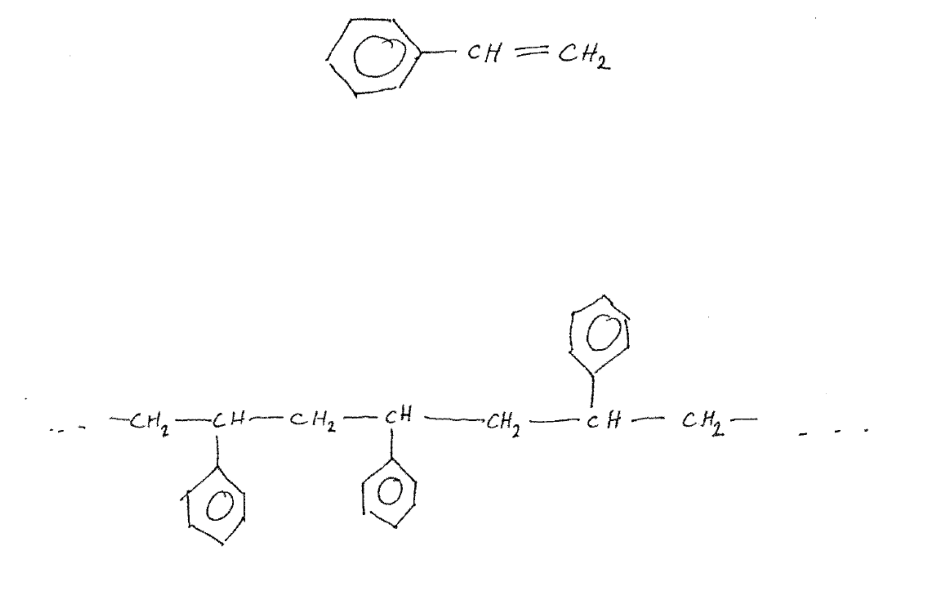}
		\caption{Styrene and polystyrene.}
		\label{fig:1}       
	\end{figure}
	\end{centering}
\begin{centering}
\begin{figure}
	\includegraphics[scale=1]{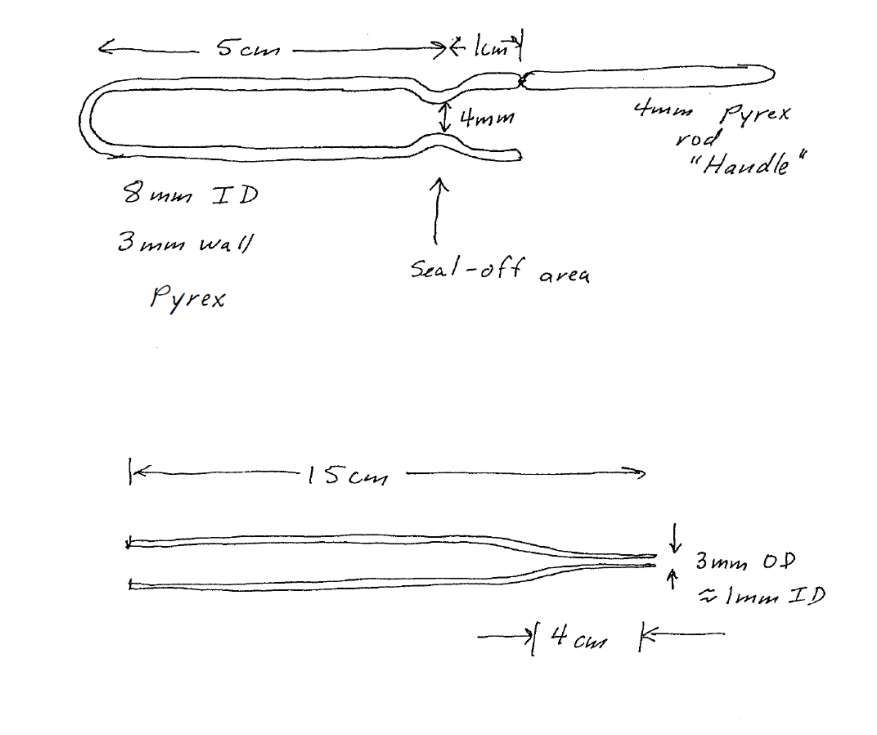}
	\caption{Reaction tube.}
	\label{fig:2}       
\end{figure}
	\end{centering}
\begin{centering}
\begin{figure}
	\includegraphics[scale=.75]{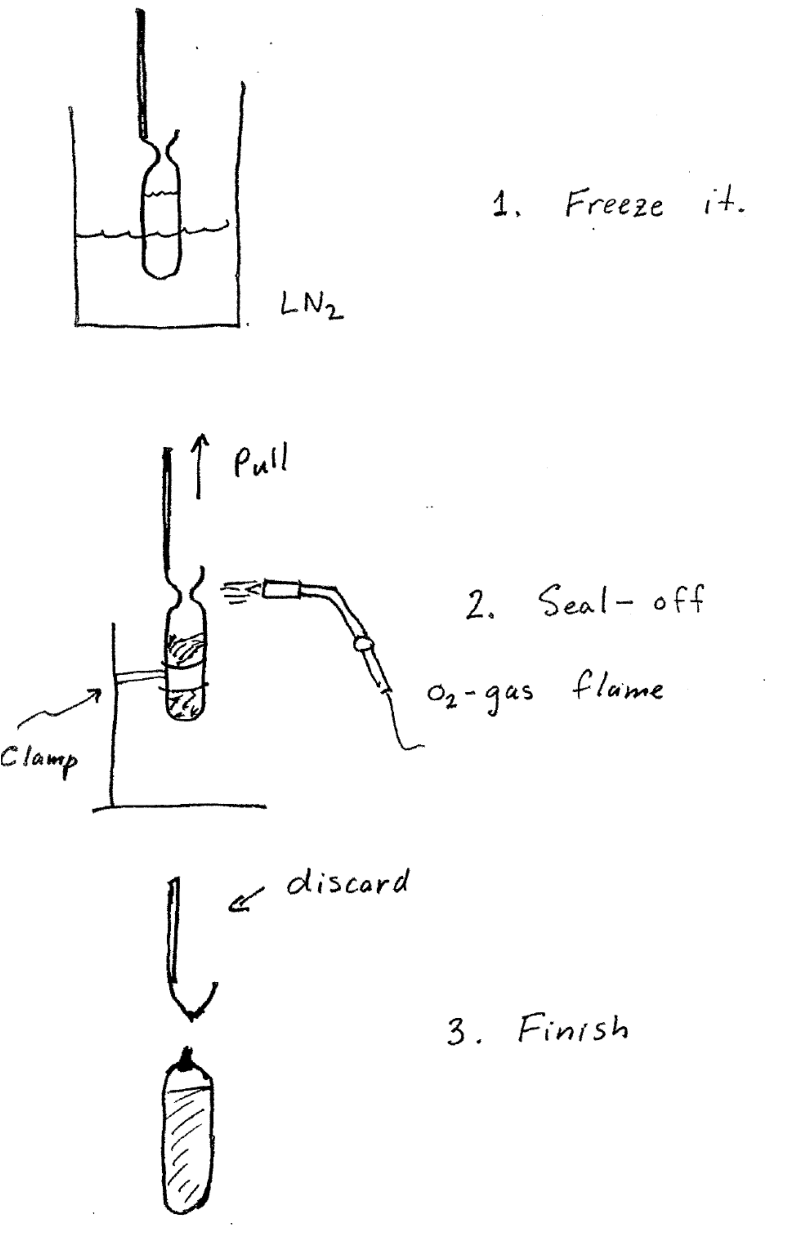}
	\caption{Handling the d-styrene.}
	\label{fig:3}       
\end{figure}
	\end{centering}
\begin{centering}
\begin{figure}
	\includegraphics[scale=1.5]{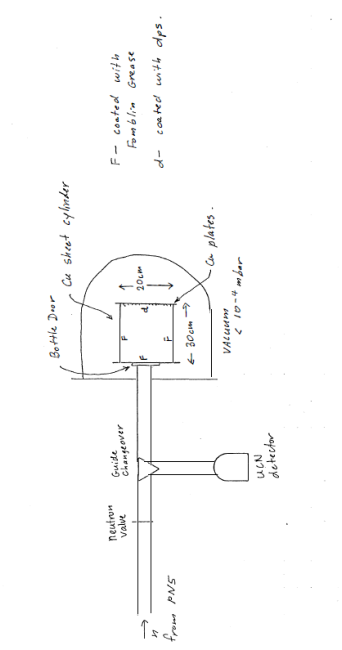}
	\caption{Lifetime measurement.}
	\label{fig:4}       
\end{figure}
	\end{centering}
\begin{centering}
\begin{figure}
	\includegraphics[scale=.75]{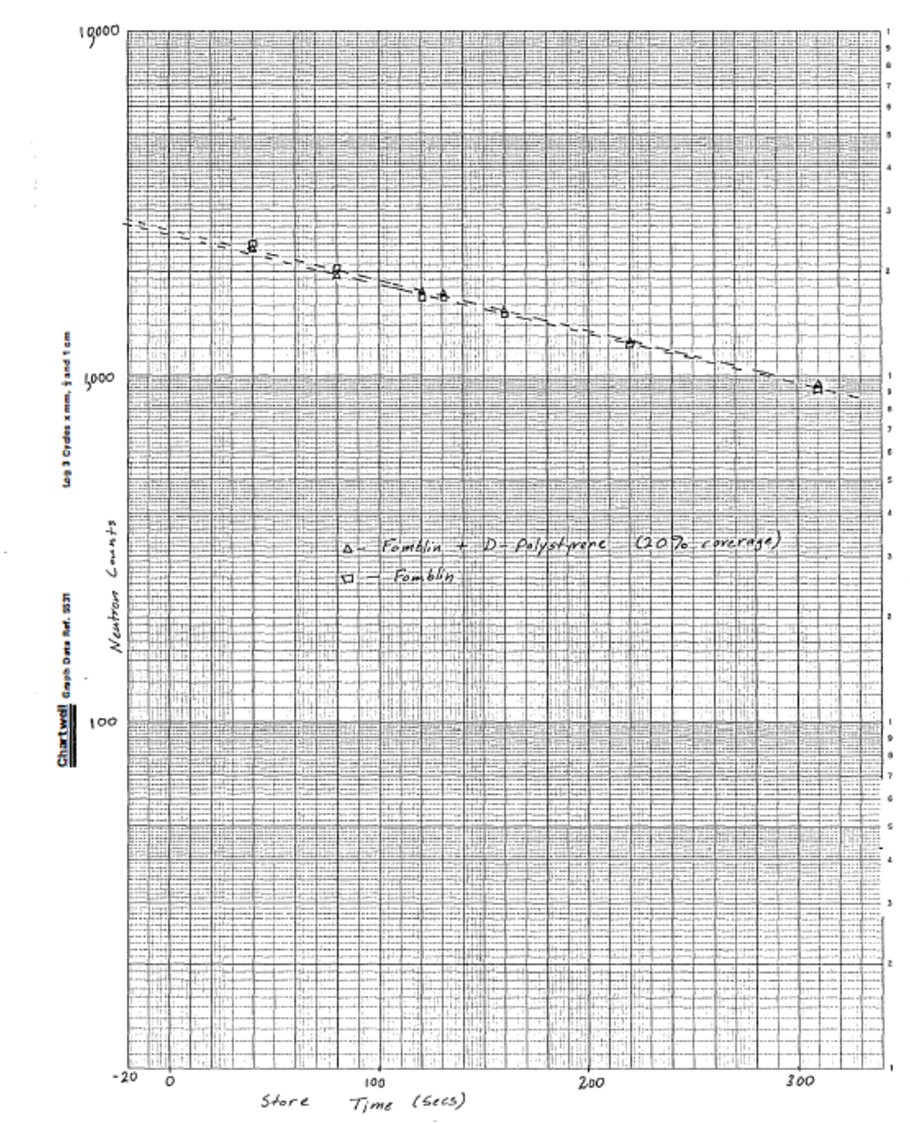}
	\caption{UCN as function of time.}
	\label{fig:5}       
\end{figure}
	\end{centering}
\begin{centering}
\begin{figure}
	\includegraphics[scale=.9]{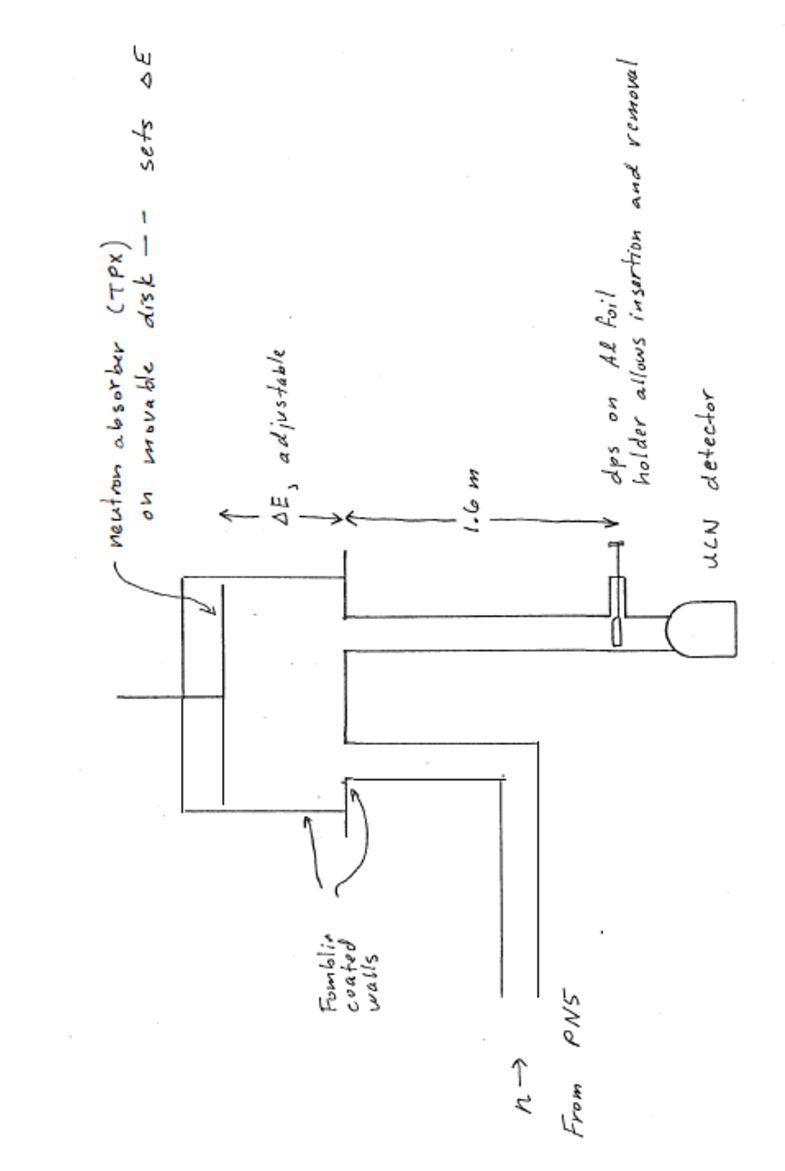}
	\caption{Energy selector.}
	\label{fig:6}       
\end{figure}
\end{centering}
\begin{centering}
\begin{figure}
	\includegraphics[scale=.75]{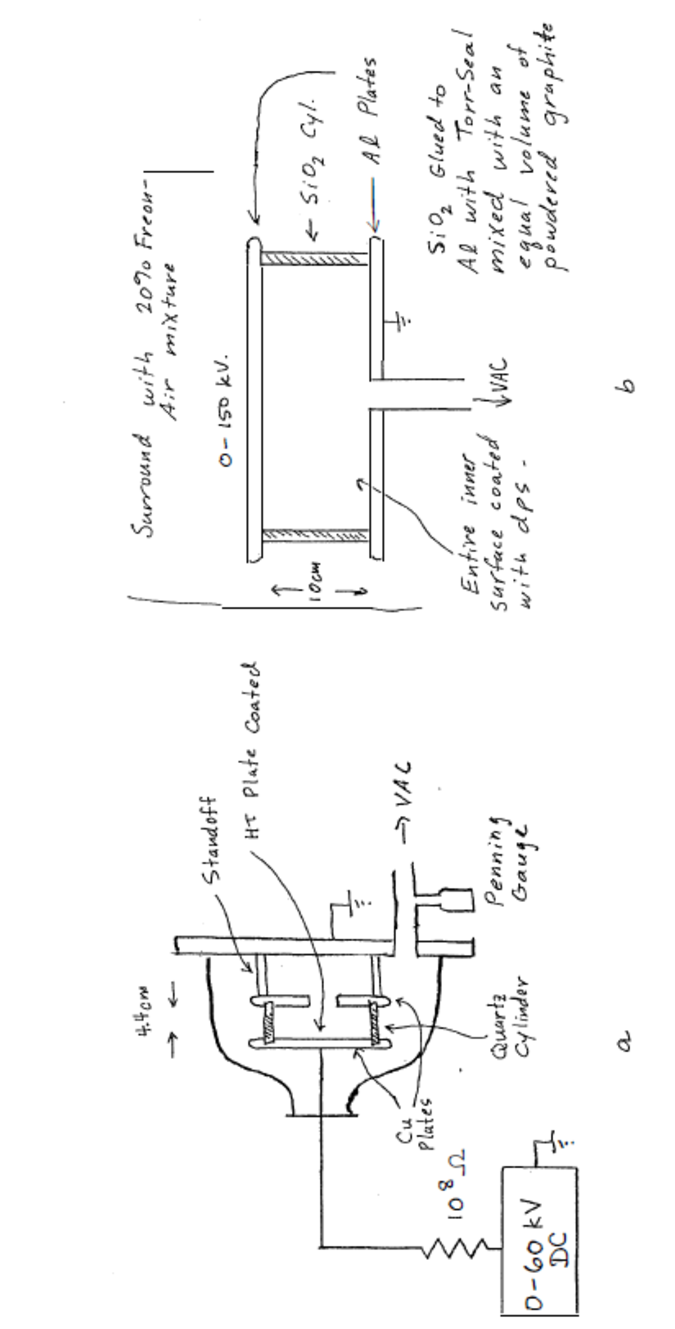}
	\caption{HT test.}
	\label{fig:7}       
\end{figure}
	\end{centering}
\begin{centering}
\begin{figure}
	\includegraphics[scale=.8]{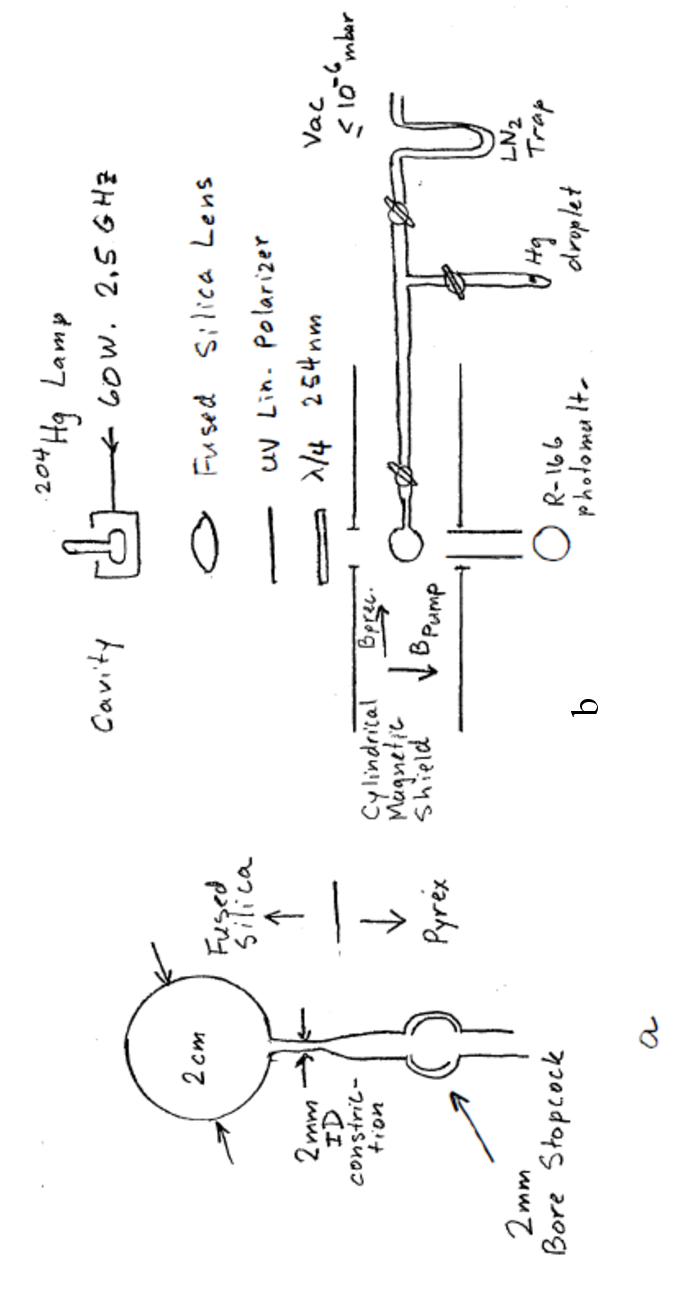}
	\caption{Hg tests cell/optical setup.}
	\label{fig:8}       
\end{figure}
	\end{centering}
\begin{centering}
\begin{figure}
	\includegraphics[scale=.8]{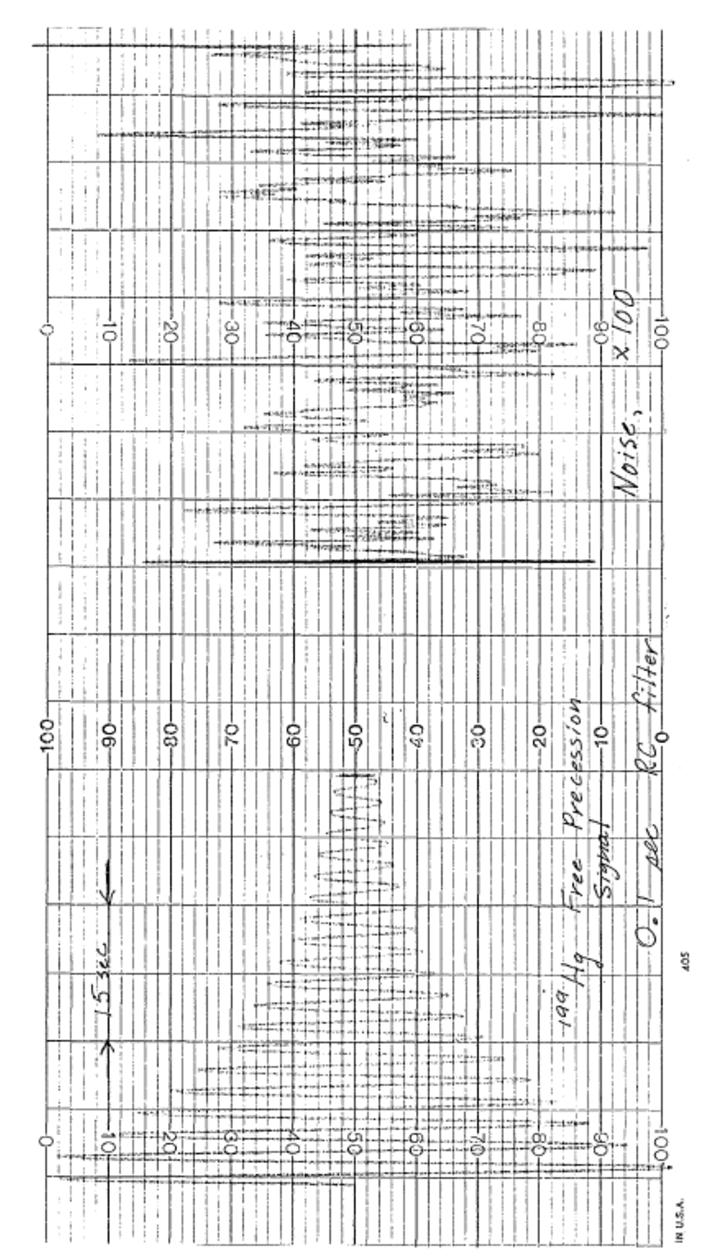}
	\caption{Hg optical pumping signal.}
	\label{fig:9}       
\end{figure}
\end{centering}
\end{document}